\documentstyle[12pt,epsf,amssymbols,here]{article}
\newcommand{\preprint}[1]{\begin{table}[t]  
            \begin{flushright}              
            \begin{large}{#1}\end{large}    
            \end{flushright}                
            \end {table}}                           
\preprint{TAUP-2335-96}
\tiny\normalsize

\begin{document}

\title{Invariant Kinematics on a One-Dimensional Lattice in Noncommutative Geometry}
\author{E. Atzmon\thanks{%
atzmon@post.tau.ac.il}  \\ 
Raymond and Beverly Sackler Faculty of Exact Sciences,\\ School of Physics
and Astronomy.\\Tel\ - Aviv University. }
\date{\today}
\maketitle

\begin{abstract}

In a one-dimensional lattice, the induced metric (from a noncommutative
geometry calculation) breaks translation invariance. This leads to some
inconsistencies among different spectator frames, in the observation of 
the hoppings of a test particle between lattice sites.
To resolve the inconsistencies between the different spectator frames, we 
replace the test particle's bare mass by an effective locally dependent 
mass. This effective mass also depends on the lattice constant - i.e. it 
is a scale dependent variable (a "running" mass).

We also develop an alternative approach based on a compensating 
potential. The induced potential between a spectator frame and the test 
particle is attractive on the average. 

We then show that the entire formalism holds for a quantum particle 
represented by a wave function, just as it applies to the mechanics of a 
classical point particle. 

\end{abstract}
\newpage

\section{Introduction}
In \cite{Atzmon,B.L.S} it was found that the distances on a one
dimensional lattice in a noncommutative geometry\footnote{Using a local 
discrete Wilson-Dirac operator} differ from their classical analog. In 
noncommutative geometry the distances on a one dimensional lattice are 
given by 
\begin{equation}
d_n=\cases{ 
            2\,a\,\sqrt{i(i+1)}~& ,  if $~n=2i~$ \cr
            2\,a\, (i+1)       ~& ,  if $~n=2i+1~$\cr
          }
\label{dist}
\end{equation}
where $i\in {\Bbb N}$, $d_{-n}=d_n$, 
and $a$ is the lattice constant, with units of length\footnote{The same 
eigenvalues for the distance operator were
obtained in 2+1 dimensions, using the Ashtekar - Lewandowski formalism 
\cite{Ashtekar}.}.

As can be seen directly, the distance to an odd $n$ point has a unit 
anomaly, whereas the distance to an even $n$ point is the geometric 
mean of the distances to the two nearby odd points. Moreover, we note
that translation invariance is broken, since $d_n-d_{n-1}\neq
d_{n+1}-d_n$. In the large $n$ limit, however, translation invariance
is restored asymptotically.\\

The fact that the distances given by (eq.\ref{dist}) thus break the
lattice's translation invariance, represents a new phenomenon of a distance's
dependence on reference frames. It follows directly from the basic
fact that all the distances in (eq.\ref{dist}) were computed from some
zero point, which the lattice sites labeling is been referred to. The
selection of one of the lattice point as the zero point, was done
without loss of generality. Therefore, {\it if several spectators are
  present on the lattice sites, with each spectator defining his
  location as the zero point\footnote{Which is logical due to the {\it
      left-right} symmetry in his own world.} in his own world, then
  due to the different sites' labeling by different spectators, the
  distance between two lattice sites becomes reference frames dependent.}

We may thus define the distance between the $i$-th and $j$-th sites,
as been labeled in some reference frame as the absolute value of the
difference between the distances i.e.

\begin{equation}
d_{i,j}=\left|\frac{j}{\left|j\right|}d_j-
  \frac{i}{\left|i\right|}d_i\right|\;\;\;\;\forall \;\;i,j \in {\Bbb Z} \label{disij}
\end{equation}

We remind the reader of the reason for the introduction of a new definition 
of distance in noncommutative geometry. It follows from the fact that the 
classical definition uses the concept of a {\it path}, i.e.
\begin{equation}
\label{eq.1}d(a,b)=\;\inf \;l_\alpha \left( a,b\right)
\end{equation}
where $l_\alpha$ is the length of the path $\alpha$ connecting 
points $a$ and $b$. The infimum is taken over all such path lengths.  
However, the concept of a path is only well defined on a smooth
manifold. One is thus forced to apply the noncommutative geometry definition,
which is more fundamental than the classical, and more general, since it 
fits all topological spaces.

This definition of a distance in noncommutative geometry is given 
by\cite{Connes}:
\begin{equation}
\label{eq.2}d(a,b)=\sup _f\left\{ \left| f(a)-f(b)\right| \;:\;f\in
A,\;\left\| \;\left[ \;D,f\;\right] \;\right\| \leq 1\right\}
\end{equation}
where $a,b\in X,\;f\in A,$\ $A$ is the algebra of functions on $X,$\ $D$ is a
Dirac operator (which is a self adjoint operator with compact resolvent)
acting in the Hilbert space $H,$\ and the norm on the r.h.s is the norm of
operators in $H$. Both definitions give the same result when the base
space is a Riemannian manifold. However the n.c.g definition has the
advantage of being applicable to discrete spaces too.\\

Noncommutative geometry, from its foundation was constructed in such
a way that it fits the concepts used in Quantum Mechanics - mainly the
passage from variables to operators. More details about the
connection to Q.M. can be found in \cite{Connes}.\\
In that context the reader should not be surprised that we were led to
the concept of reference\,-\,frames\,-\,dependent distances. Not with
standing its
novelty, what can be more natural in Q.M. than an influence of
the observer on the results? We believe that a similar phenomenon will be
revealed in a full theory of Quantum Gravity when this will be found.

In the following, we analyze some of the possible physical implications of
the above result.

\section{The model}

Assume a one dimensional lattice with lattice constant\ $a$. On the lattice 
there are a {\it spectator} and a {\it test particle (t.p.)}. We assume that 
the spectator sticks to one site, which serves as the zero reference point 
in the spectator's system, from which distances are measured on the lattice. 
The t.p. has a mass $\mu $ and is situated at a site $i$. The dynamics of the 
t.p. on the lattice are characterized by a hopping motion from site to site on 
the lattice, where for each jump to occur there is some probability. Such  
dynamics are particularly appropriate in a scenario in which the system is a  
quantum system, in analogy with the motion of a particle between discrete 
energy levels. Note that the dynamics on the lattice can not be "smooth", i.e. 
the t.p. can jump between two lattice sites without "passing" through the 
intermediate sites, as in jumps between Bohr's orbits for atomic electrons.
Essentially, such hopping dynamics are known from solid state physics 
(e.g. Mott hopping conductance, also connected to percolation theory 
\cite{Mott}). Quantum behavior is very much in the spirit of dynamics in a 
noncommutative geometry.

Two questions now arise. First, how do we treat {\it time}? The second question 
relates to the {\it probabilities} for the test particle's hops -- how do we 
go about evaluating them?

\subsection{The issue of time}

As we shall see, there are three possible answers (and perhaps even more) 
for the first question. The first possible answer would claim that there 
might be no need at all for an answer, since one can deal with probabilities 
without involving time dependence, in the same way as one deals with the 
probabilities in dice, namely, without caring about the rate at which the 
dice are thrown. Thus, in this solution, the jumps' rate does not affect 
the probabilities. However, having in mind physical models and not just 
mathematical ones, we believe that the question cannot be eliminated. 
The second possible answer is that time is an extra dimension, as is 
usually done. Thus the total spacetime is ${\Bbb Z}\otimes {\Bbb R}$ or 
${\Bbb Z}\otimes {\Bbb Z}$, depending on whether the time is continuous 
or discrete, respectively. But, since the results for the
distances on a lattice which we use here are taken only for the one-
dimensional case, and the distances on those two-dimensional spaces could be 
different from what we use, we prefer not to adopt this answer. 
Thus, although the second answer might be the correct one physically,  
we select for our purposes the following third answer, which enables us 
also to apply our result (\ref{dist}) for the distances. 
This third answer consists in using our quantum hops to define and 
measure time intervals, in the spectator system. {\it To 
measure time, we count the t.p. jumps}. In this context, time is not an
extra dimension, but rather a parameterization defined by counting, similar to 
the counting of the vibrations of the quartz in a digital watch.  
Although some vibrations might have longer periods (as measured by counting 
the vibrations of a more accurate watch) time can still be defined
as the result of an exact counting of the vibrations.
\footnote{As one can see, with zero velocity not being allowed for a
  particle in the third option, one regains the Q.M. concept that the
  location and the momentum can not be simultaneously defined, as well
  as the classical concept that time can not be stopped.}

\subsection{The probabilities issue}

To determine the probabilities for the t.p.'s hops on the lattice, we first 
examine the propagator of a free boson on a translation-invariant space:  
\ $\frac 1{p^2+\mu ^2}$. The Fourier transform of this propagator is 
essentially the correlation function between different locations in the 
coordinate space. It seems reasonable to take the correlation function 
between two different sites as proportional to the probability of a jump 
between them: the more they are correlated, the stronger the chances 
for the t.p. to make that jump. Working with a one-dimensional lattice, 
we take the one-dimensional Fourier transform of the propagator. 
The probabilities will thereby be of the form: 
\begin{equation}
P_{i,i+j}=\frac{e^{-\beta \left| x_i\;-\;x_{i+j}\right| }}{\beta f_i\left(
\beta \right)} = \frac{e^{- \frac{\beta d_{i,j}}{a}}}{\beta f_i\left(
\beta \right)}   \label{prob1}
\end{equation}
where $\beta =\mu a\;$($\mu $ - is the t.p. mass, and $a$ is the lattice
constant), $i$ and $i+j$ are the initial and the final sites of the t.p.
relative to the spectator, $x_n=\frac n{\left| n\right| }\frac{d_n}a$, and $
f_i\left( \beta \right) $ is the normalization of the probabilities. 
Remembering that the lattice translation invariance is broken and only reached 
asymptotically - we take the above expression for the probabilities only as 
a first guess, a form that they should roughly follow. The normalization is
given by the following formula: 
\begin{equation}
f_i\left( \beta \right) =\frac 1\beta \sum_{j=-\infty }^\infty e^{-\beta
\left| x_i\;-\;x_{i+j}\right| }  \label{norm1}
\end{equation}
Note that for $i=0$ the t.p. is initially situated at the same site as the
spectator.
 
As can be immediately seen, if $\beta $ is a constant, the hopping 
probabilities for the t.p. as seen in the spectator's frame differ from 
those evaluated in the t.p. frame. This is a direct outcome
of the fact that $\left| x_i-x_{i+j}\right| \neq \left| x_j\right| $ where
the l.h.s is the distance as observed in the spectator frame and the r.h.s is
the distance as observed in the t.p. frame. In the following section, we  
discuss the meaning of this incompatibility between the probabilities
in the two systems, and ways of resolving this difficulty.

\section{Resolving the probability paradox}

\subsection{Further aspects of the incompatibility between the probabilities in
the spectator and the t.p. frames.}

A closer look at the distances involved will reveal that for a t.p. initially 
situated at an even site $i=2l$, the most probable jump, as evaluated in the 
spectator frame is to the site $2l-1$ (i.e. $j=-1$). If the t.p. was initially 
located at the site $i=2l-1$, the most probable jump, as evaluated in the 
spectator frame is to $2l$ (i.e. $j=+1$). However in the t.p. own frame, 
there is no difference in the probabilities for jumps to the left or to the 
right.

Another anomaly which is an outcome of the above incompatibility can be 
seen when inspecting distances between sites whose indices differ by
two units. Though the distance between two neighboring odd points is always 
$2a$ (except for $i_1=-1 , i_2=1$), the 
distances between two neighboring even points form a decreasing series, 
tending to $2a$, when $n\rightarrow\infty $. In the spectator frame, the 
effective probability for the test particle's motion will thus be larger
in an outward direction (relative to the spectator), i.e. the t.p.
{\it drifts away}. In the t.p. frame, however, as there is no 
difference between its two sides, the t.p. should just stay in its initial 
position (on the average). Note, moreover, that different spectators - located each 
at a different lattice site (i.e. with different origins in their 
respective frames) - will observe different probability distributions.

Though the probability space in the t.p. frame is a commutative space 
(in the sense that when defining $P_j$ \ as the probability of jumping $j$ 
sites to the right, then $P_kP_j=P_jP_k$), it is noncommutative in the 
spectator frame, where $P_kP_j\neq P_jP_k$. This 
point summarizes the inconsistencies between the two frames.

Since the probabilities are thus not the same, whether between spectator 
and t.p. frames, or between two spectator frames, applying our "third option" 
as defined in section 2.1, we observe that {\it the time intervals are 
not synchronized}. We thus set upon removing the anomalies by an 
appropriate corrective mechanism.

\subsection{Resolving the incompatibility: the frame-dependent mass solution}

All above anomalies were obtained under the assumption that the mass - 
$\mu $, of the t.p. is the same in both the spectator and the t.p. frames. 
Our solution is to abandon the assumption that the t.p. mass is a global 
parameter and assume instead that it is a local variable. In the following 
we demonstrate that this is a complete resolution of our paradoxes.

We require the probabilities $P_{i,i+j}$ and $P_{0,j}$ to be equal in both 
spectator and t.p. frames, i.e. using localized masses, we restore 
translation invariance for the probability of hopping $j-$sites. 

\begin{equation}
\forall \;i,j\;\in {\Bbb Z\;\;};{\;\;\;\;\;\;\;\;\;}P_{i,i+j}=\frac{%
e^{-a\mu _{i,i+j}\left| x_{i+j}-x_i\right| }}{a\mu _{i,i+j}f\left( a\mu
_0\right) }=\frac{e^{-a\mu _0\left| x_j\right| }}{a\mu _0f\left( a\mu
_0\right) }=P_{0,j}  \label{prob2}
\end{equation}
where $\mu _{i,i+j}$ is the mass of the t.p. as observed in the spectator 
frame, while hopping from the $i$ site to the $i+j$ site,
and $\mu _0$ is the mass in the t.p. system. The normalization constant $
f\left( a\mu _0\right) $ is defined as:
\begin{equation}
f\left( a\mu _0\right) =\sum_{j=-\infty }^\infty \frac{e^{-a\mu _0\left|
x_j\right| }}{a\mu _0}=\frac 2{a\mu _0}\left( \frac 12+\frac{e^{-2a\mu _0}}{%
1-e^{-2a\mu _0}}+\sum_{j=1}^\infty e^{-2a\mu _0\sqrt{j\left( j+1\right) }
}\right)   \label{norm2}
\end{equation}
where we are using (\ref{dist})[see figure 1 in the Appendix], thus 
normalizing the probabilities (i.e. $
\;\sum_{j=-\infty }^\infty P_{0,j}=\sum_{j=-\infty }^\infty P_{i,i+j}=1$).

From (\ref{prob2}) we obtain the following relation between $\mu _{i,i+j}$ and 
$\mu _0$:
\begin{equation}
a\left( \mu _0\left| x_j\right| -\mu _{i,i+j}\left| x_{i+j}-x_i\right|
\right) =\ln \left( \frac {\mu _{i,i+j}}{\mu _0} \right)  \label{mass2}
\end{equation}

As one can see, equation(\ref{mass2}) can not be analytically solved for $
\mu _{i,i+j}$, but if one assumes for simplicity that:
\begin{equation}
a\mu _{i,i+j}\left| x_{i+j}-x_i\right| \ll 1\;;\;\;\;\;\;a\mu _0\left|
x_j\right| \ll 1  \label{limit}
\end{equation}
one finds from (\ref{prob2}) that the mass as seen by the spectator is 
given by
\begin{equation}
\mu _{i,i+j}\simeq \frac{\mu _0}{1+a\mu _0\left[ \left| x_{i+j}-x_i\right|
-\left| x_j\right| \right] }\approx \mu _0\left[ 1-a\mu _0\left( \left|
x_{i+j}-x_i\right| -\left| x_j\right| \right) \right]  \label{mass}
\end{equation}

We emphasize that if one assigns Planck-length value $l_p$\ to the lattice 
constant $a$ (i.e. $\sqrt{\frac{\hbar G}{c^3}}$\ ), then the limit
(\ref{limit}) should be interpreted as the low mass limit $\mu \ll m_p$\ ,
where $m_p$ is a Planck-mass (i.e. $l_p^{-1}$). Thus, the mass formula (\ref
{mass}) should be seen as semi-classical.

By setting the probabilities to be equal in the two frames, we are essentially
synchronizing the relevant times - applying our definition of time intervals, 
based on counting jumps. The price, however, has been paid by the 
transformation of the mass $\mu_0$ into a local and dynamical variable - 
$\mu_{i,i+j}$ (local and dynamical because of the $i$ and the $j$ dependence 
respectively). 

There is a very rough analogy with the emergence of the Doppler effect in 
Special Relativity. However, whereas in the Doppler effect the frequency 
of a wave emitted by a system moving towards the spectator is blue-shifted, 
i.e. always increases in the spectator frame, in our case, the t.p. mass as 
observed in the spectator frame $\mu_{i,i+j}$ can be smaller than $\mu _0$ ,
even when the t.p. jump is directed towards the spectator.
The fact that the physical mass $\mu_{i,i+j}$ depends on the lattice constant 
$a$, aside from its dependence on $\mu_0$ and on the location, 
suggests that $\mu_{i,i+j}$ depends on the scale. This behavior allows us 
to think of the mass as a ``running mass'', as happens in the
context of the renormalization procedure in field theory (see figure 2 in
the Appendix).

Adopting the above concept of a local mass, we have thus achieved full 
agreement between the spectator and t.p. probability values. There is thus 
also a synchronization of time measurements between the frames, and there is 
no preferred direction on the lattice.

\subsection{The potential approach}

Instead of modifying the kinematics, by constraining the mass to be a 
local variable, in our effort at restoring covariant dynamics for the t.p., 
we could use an alternative, "dynamical" approach, by introducing 
potentials. With the distances on the lattice not being translation 
invariant, the propagator of a boson particle situated at the $i-th$ site 
(in the spectator frame) could be written in terms of a potential $V_{i}$,, 
\begin{equation}
\left[ \;p^2+\mu _0^2-V_i\left( \left| x_{i+j}-x_i\right| \right) \;\right]
^{-1}  \label{propagator}
\end{equation}
where $V_i\left( \left| x_{i+j}-x_i\right| \right) $ is induced by the 
metric defined in (\ref{dist}) and (\ref{disij}).\ As we demonstrated, 
this is, on the average, a repulsive force,

\begin{equation}
F_{avg}\equiv -\left( \frac{\partial V}{\partial x}\right) _{avg}>0
\label{force}
\end{equation}

To prevent this outward drift of the t.p. in the spectator frame, 
and to establish a correspondence between the t.p.'s propagation in all 
spectator frames, we then postulate {\it an induced attractive potential   
$ V_i^{ind}\left( \left| x_{i+j}-x_i\right| \right) $\ {\it which will 
exactly cancel} $V_i\left( \left| x_{i+j}-x_i\right| \right) $. This is 
somewhat reminiscent of Einstein's 1917 introduction of the cosmological 
constant, to stabilize the universal geometry. The propagator of the t.p. 
in all spectator frames are now the same as for a free particle 
$\left[ p^2+\mu^2\right] ^{-1}$. The existence of the induced potential, 
in some aspects, is also similar to the Lentz law in classical 
electromagnetism.

One can identify the modified mass in (\ref{prob2}) with the following
effective mass, defined as: 
\begin{equation}
\mu _{i,i+j}^{eff}=\sqrt{\mu _0^2+V_i^{ind}\left( \left| x_{i+j}-x_i\right|
\right) }=\sqrt{\mu _0^2-V_i\left( \left| x_{i+j}-x_i\right| \right) }
\label{effmass}
\end{equation}
In other words, one can deduce from $\mu _{i,i+j}$ the value of the induced
potential between the t.p. and the spectator (in the low mass limit 
(\ref{limit}) one can use (\ref{mass}) and (\ref{effmass})).} Summarizing 
this approach, we note that though the distances break translation invariance  
and while the mass preserves its global definition, the hopping 
probabilities -- and the related beats of the clock -- have become translation 
invariant.\\

{\Large Summary}\\
We have shown that the induced metric of a one-dimensional lattice, 
(from a noncommutative geometry calculation) breaks the lattice translation invariance and therefore also becomes reference-frame dependent, possibly leading to some paradoxes concerning the fundamental issue of general covariance. We have suggested a solution of the paradoxes, essentially at the probabilistic level - which in our opinion is the most important aspect, at least from the measurement aspects of Quantum Mechanics. It was done by either allowing an effective locally dependent mass which thus becomes a ``running mass'', or by introducing a compensating potential. Both ways lead to a commutative probabilistic space which is general covariant.\\   

{\large Acknowledgment}\\
I am most greatful to Prof. Yuval Ne'eman for the enlightening
comments. I would like also to thank the anonymous referee for his
constructive suggestions.

\newpage  

{\LARGE Appendix}\\ 

{\Large A. The test particle as a wave function}\\
Up to this point, we have treated the t.p. as a classical point particle. 
In the following, we go over to quantum mechanics - i.e. the t.p. is now 
a quantum particle, represented by a wave function - 
$\varphi \left( k\right) $\ which is defined over the entire lattice.

Let $\varphi \in \cal{L}$$^2$ (i.e. a square integrable function) be 
normalized so that $\left\|
\varphi \right\| =1$.\ Let $\left( \psi _j\right) _{j\in {\Bbb Z}}\;$be a
complete set of orthogonal states which span the Hilbert space -
${\Bbb H\;}$%
which is defined over the lattice. 
The Hilbert space basis 
$\left( \psi _j\left( k\right) \right)_{j,k\;\in {\Bbb Z}}$\ is defined 
as follows:
\begin{equation}
\psi _j\left( k\right) =\delta _{j,k}\label{delta}
\end{equation}
where the $j$ -index stands for a state in the Hilbert space, while the
$k$ -index stands for the lattice site.  
Any $\varphi $ can now be represented by the following sum:
\begin{equation}
\varphi =\sum_{j=-\infty }^\infty \alpha _j\psi _j\label{cigma} 
\end{equation}
where the $\alpha _j$ -s' are complex numbers. Since $\left\| \varphi \right\|
^2=1$ it follows that $\sum_{j=-\infty }^\infty \left| \alpha _j\right| ^2=1$
.

As in Q.M. the $\left| \alpha _j\right| ^2$ represent the probability for 
the particle to be found in a measurement at the $j$-site of the lattice. 
One can thus treat the $\left\| \varphi \right\| ^2$ as a distribution, for  
which the $\left| \alpha _j\right| ^2$ -s' form a probability space.

Let $\left\| \varphi \right\| ^2 _i$ be a distribution centered at the $i$ 
site of the lattice. We define $ (\alpha _j)_i $ as the coefficients in 
(\ref{cigma}) for the $\left\| \varphi
\right\| ^2_i$. We observe the similarity between 
$\left|\alpha_j\right| ^2 _i$ 
and $P_{i,i+j}$; both define the probability for the t.p. to be initially 
at the $i$ - site and after a measurement, in the $(i+j)$ - site. We have 
thus essentially showed that the t.p. can also be represented by a wave 
function, with the effective mass (or the induced potential) having the 
same features as before.\\ \\

{\large B. In the following we exemplify some of the formulae found in the article.}

\vspace{10mm}
\epsfxsize=11truecm
\centerline{\epsfbox{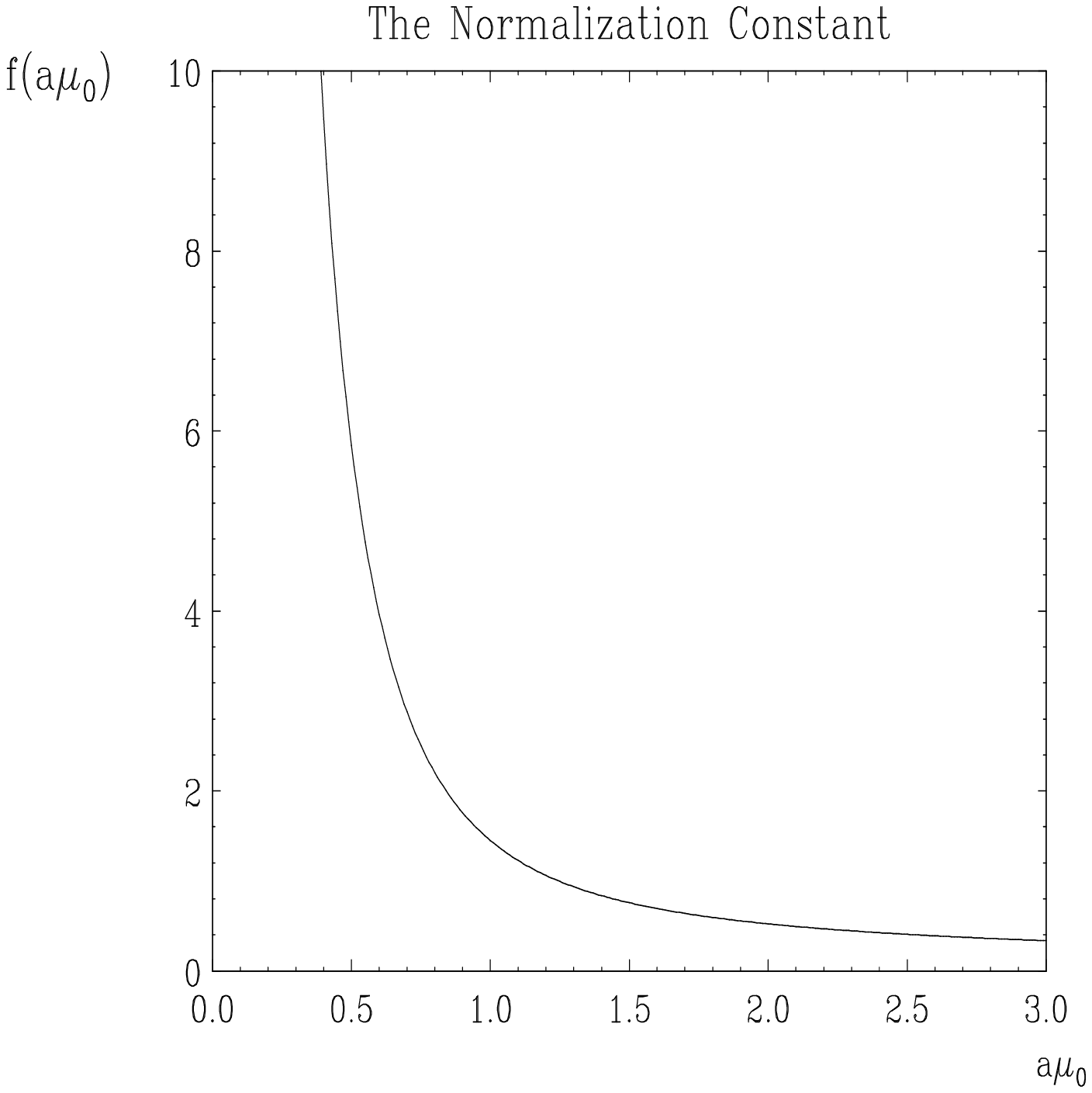}}

\vspace{2mm}
\centerline{\parbox{11truecm}{Figure 1.{\footnotesize The dependence of 
the normalization constant given in
(\ref{norm2}) on $a \mu_0$ (i.e. on the product of the lattice
 constant by the ``bare" mass).}}}
\vspace{5mm}

Figures 2a - 2e reproduce some
examples of the running mass in (\ref{mass2}). The $i$ and the $j$
indices represent the initial and the hopped-to sites of the
test particle, respectively.
$\mu_0$ is the t.p. mass in its own frame, and $a$ is the lattice
constant.

\begin{figure}[H]
\epsfxsize=6.5truein
\centerline{\epsffile{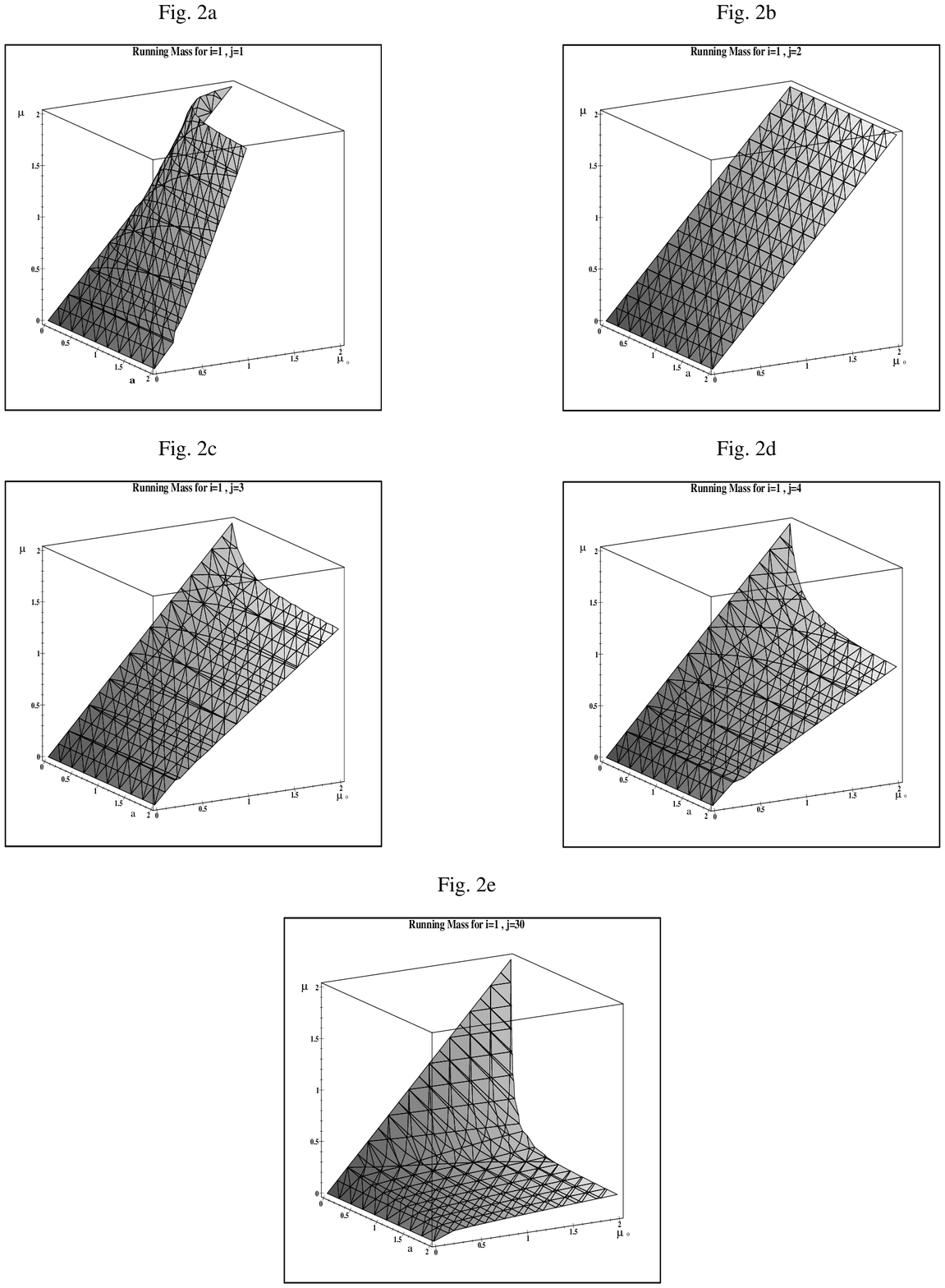}}
\label{fig1}
\end{figure}

\end{document}